\documentclass[11pt,twoside]{article}
\usepackage{./asp2014}

\newcommand{\hi}{\mbox{H{\sc i}}}
\newcommand{\msol}{\rm M$_\odot$}
\newcommand{\kms}{km s$^{-1}$}

\aspSuppressVolSlug
\resetcounters

\begin{document}

\title{Cosmic Web Research with KAT-7, MeerKAT \& FAST}
\author{Claude Carignan$^{1,2}$ 
\affil{$^1$Department of Astronomy, University of Cape Town, Private Bag X3, Rondebosch 7701, South Africa; \email{ccarignan@ast.uct.ac.za}}
\affil{$^2$Observatoire d'Astrophysique de l'Universit\'e de Ouagadougou (ODAUO), Burkina Faso}
}

% This section is for ADS Processing.  There must be one line per author.
\paperauthor{Claude Carignan}{ccarignan@ast.uct.ac.za}{}{University of Cape Town}{Astronomy Department}{Rondebosch}{Western Cape}{7701}{South Africa}

\begin{abstract}
The seven-dish KAT-7 array was built as an engineering test-bed for the 64-dish Karoo Array Telescope, known as MeerKAT, which is the South African precursor of the Square Kilometre Array (SKA). KAT-7 and MeerKAT are located close to the South African SKA core site in the Northern Cape's Karoo desert region. Construction of the KAT-7 array was completed in December 2010. The short baselines (26 to 185 m) and low system temperature (T$_{\rm sys} \sim$ 26 K) of the telescope make it very sensitive to large-scale, low-surface-brightness emission, which is one of the \hi\ science drivers for MeerKAT and one of the future strengths of FAST.  While the main purpose of KAT-7 was to test technical solutions for MeerKAT and the SKA, scientific targets were also observed during commissioning to test the \hi\ line mode and the first results obtained are presented.
A description of MeerKAT and an update on its construction is also given. Early science should start in mid-2016 with a partial array and the full array should be completed at the end of 2017. For cosmic-web research (detection of low column density \hi), a future combination of data from FAST and MeerKAT should allow to explore the unknown territory of $< 10^{18}$ cm$^{-2}$ surface densities and the possible connection between spiral galaxies and the surrounding cosmic web.
\end{abstract}

\section{Introduction}

In the $\Lambda$CDM cosmological model, galaxies and large scale structures form hierarchically with the baryonic matter (stars and gas) following the DM density distribution. With radio telescopes reaching lower and lower \hi\ column densities below $10^{18}$ cm$^{-2}$, it should soon be possible to link the \hi\ in the outer parts of galaxies to the surrounding cosmic web seen in numerical simulations \citep[see e.g. Millennium:][]{spr05}. Already, very deep \hi\ surveys with Westerbork \citep[see e.g. HALOGAS:][]{hea11} or the GBT \citep[see e.g.][]{wol15} start detecting \hi\ in the outer parts of galaxies' halos and their surroundings down to densities of a few $\times 10^{17}$ cm$^{-2}$ at low spatial resolution. In the SKA era, SKA1-MID should reach, in an 8 hours observation, densities below $10^{18}$ cm$^{-2}$ with a $\sim$30\arcsec\ beam and below $10^{17}$ cm$^{-2}$ with $\sim$100\arcsec\ resolution \citep{pop15}.

What do KAT-7 (Fig. \ref{fig:kat7}) and MeerKAT (Fig. \ref{fig:meerkat}) have in common with FAST for cosmic web research ? 
KAT-7 and FAST have the same spatial resolution $\sim$3\arcmin\ at \hi\ but FAST will have $\sim$90 times more collecting area. 
On the other hand, while MeerKAT will have $\sim$8 times less collecting area than FAST,  it will have $\sim$6 times better spatial resolution $\sim$30\arcsec. As we will see later, soon the perfect combination for cosmic web studies could be to combine the data from FAST for the sensitivity and from MeerKAT for the spatial resolution.

\articlefigure[width=.9\textwidth]{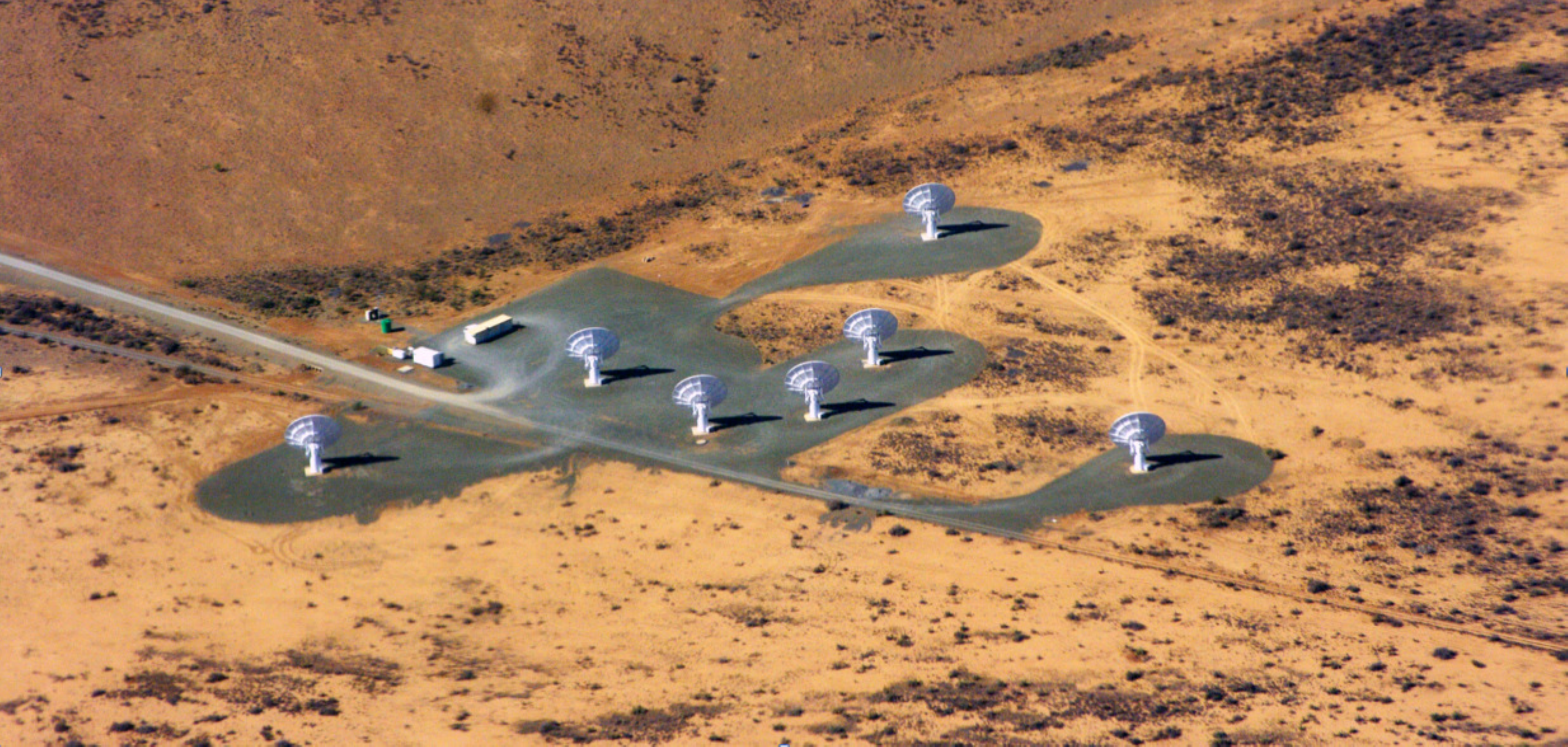}{fig:kat7}{The pathfinder KAT-7 in the Karoo desert in South Africa.}

%\articlefigure[width=.6\textwidth]{MeerKAT.eps}{fig:meerkat}{The first MeerKAT antenna on the SKA site.}

\articlefigure[width=.7\textwidth]{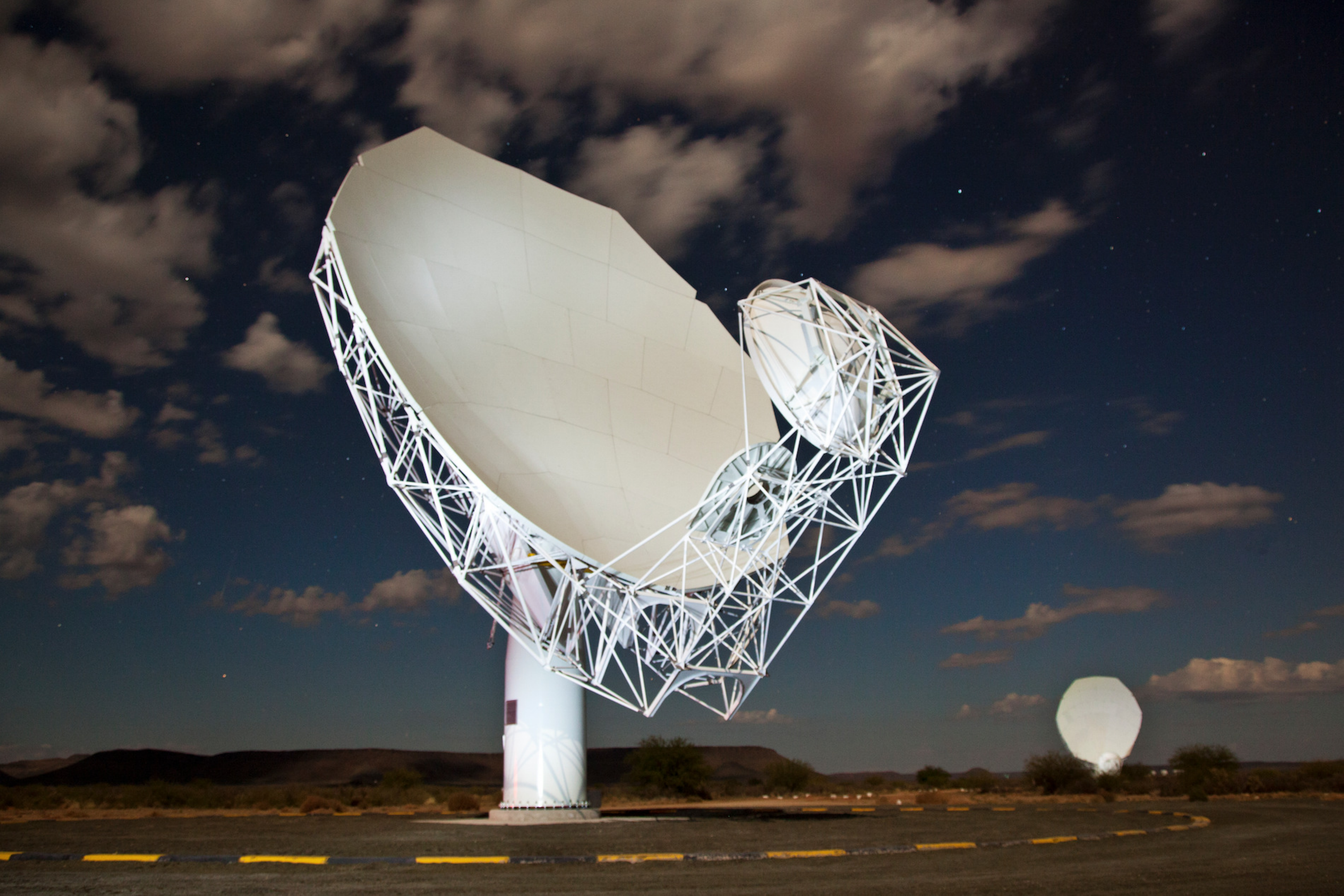}{fig:meerkat}{A MeerKAT antenna with its offset Gregorian feed.}

\section{SKA SA Radio Astronomy Roadmap}

South Africa has a well defined roadmap up to the completion of the second phase of the SKA around 2030. It can be summarised as follows:
\begin{itemize}
\item {\bf Phase 1:} the first step was the construction of the pathfinder KAT-7 with 7 $\times$ 12 m antennas. The telescope was completed in December 2010 and is in operation since then. Some of the results from the completed \hi\ projects will be described below.
\item {\bf Phase 2:} the second step is the construction of the SKA precursor MeerKAT, which will have 64 $\times$ 13.5 m antennas with maximum baselines of 8 km. This project is fully funded by the South African government and should be completed toward the end of 2017, with early science starting in 2016.
\item {\bf Phase 3:} the third step will be to merge MeerKAT with SKA1-MID. After the rebaselining exercice, this will mean merging the 64 MeerKAT antennas with 133 new 15 m SKA antennas with maximum baselines $\sim$150 km. SKA1-MID will be constructed between 2018 and 2023, with early science starting in 2020.
\item {\bf Phase 4:} the last phase, which still have to go through a final design period (2018-2021), will be built between 2023 and 2030. It will have of the order of $\sim$2500 antennas to reach the square kilometer collecting area.
\end{itemize}

\section{Early \hi\ results with KAT-7}

Figure \ref{fig:n3109} shows the first \hi\ observation of the Magellanic-type spiral NGC 3109 \citep{car13}. The short baselines and low system temperature of the telescope make it sensitive to 
large scale low surface brightness emission.  The new observations with KAT--7 allowed the measurement of the rotation curve of NGC 3109 out  to 32\arcmin , 
doubling the angular extent of existing measurements. A total \hi\ mass of 4.6 $\times 10^8$ \msol\
was derived, 40\% more than what was detected by previous VLA observations. 

\articlefigure[width=.6\textwidth]{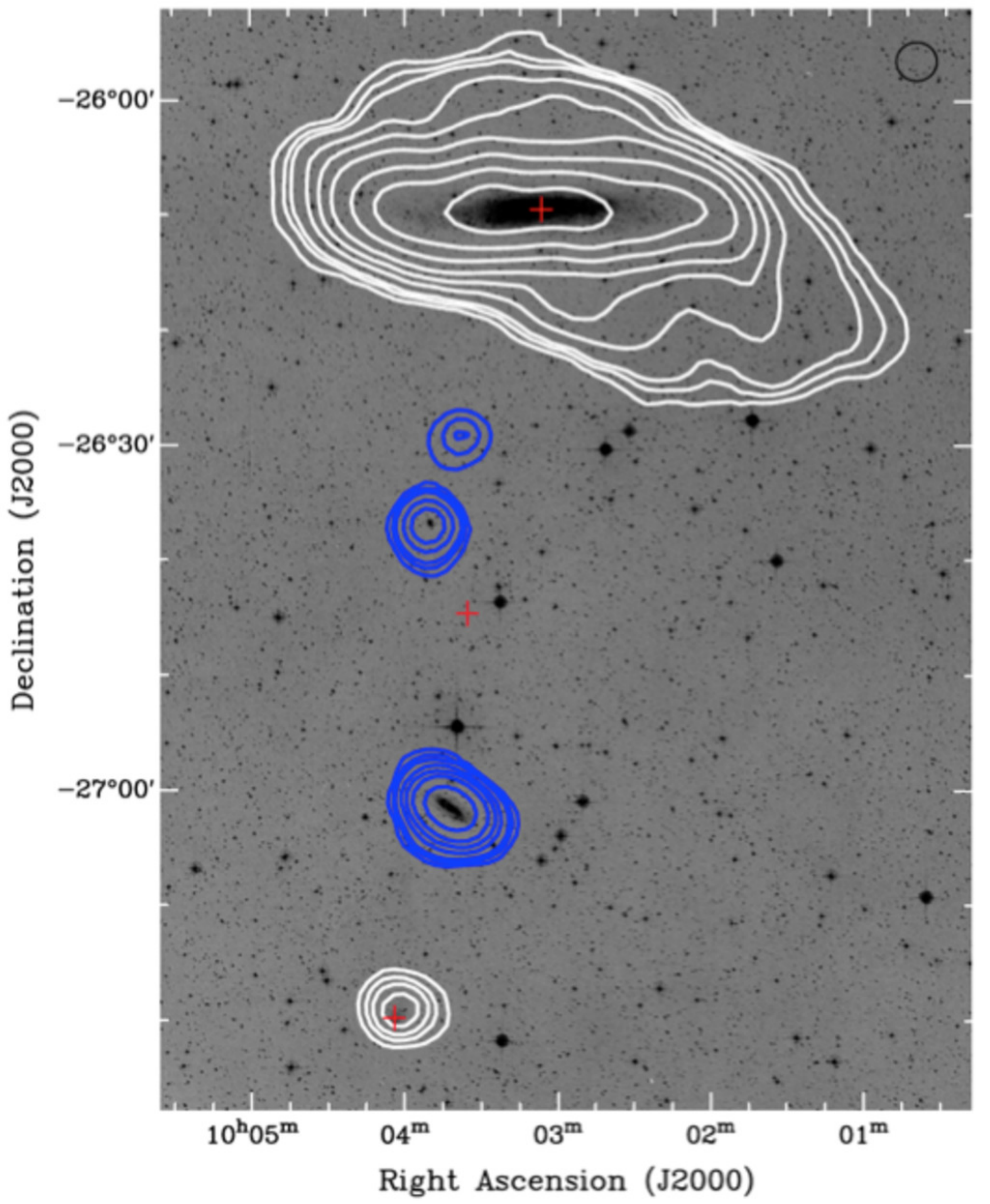}{fig:n3109}{\hi\ distribution of NGC 3109 (top) and of the Antlia dwarf (bottom). The blue countours are background galaxies.}

Figure \ref{fig:n253} shows the second \hi\ observation of  the Sculptor Group starburst spiral galaxy NGC 253 \citep{luc15}.
The KAT-7 observations detected 33\% more flux than previous VLA observations, mainly in the outer parts and in the halo for a total \hi\ mass of $2.1 \times 10^{9}$ M$_{\odot}$. A significant fraction of the  \hi\ can be found away from the plane out to projected distances of $\sim$9-10 kpc in the centre and 13-14 kpc at the edge of the disk.
A  novel technique, based on interactive profile fitting, was used to separate the main disk gas from the {\it anomalous} (halo) gas. Full 3D models were generated to check the consistency of the adopted kinematical parameters. The rotation curve  (RC) derived for the \hi\ disk confirms that it is declining in the outer parts, as seen in previous optical Fabry-Perot measurements. As for the anomalous component, its RC has a very shallow gradient in the inner parts and turns over at the same radius as the disk, kinematically lagging by $\sim$100 \kms. The kinematics of the observed extra planar gas are compatible with an {\it outflow} due to the central starburst and galactic fountains in the outer parts. However, the gas kinematics show no evidence for {\it inflow}. Analysis of the near-IR WISE data, shows clearly that the star formation rate (SFR) is compatible with the starburst nature of NGC 253.

\articlefigure[width=.7\textwidth]{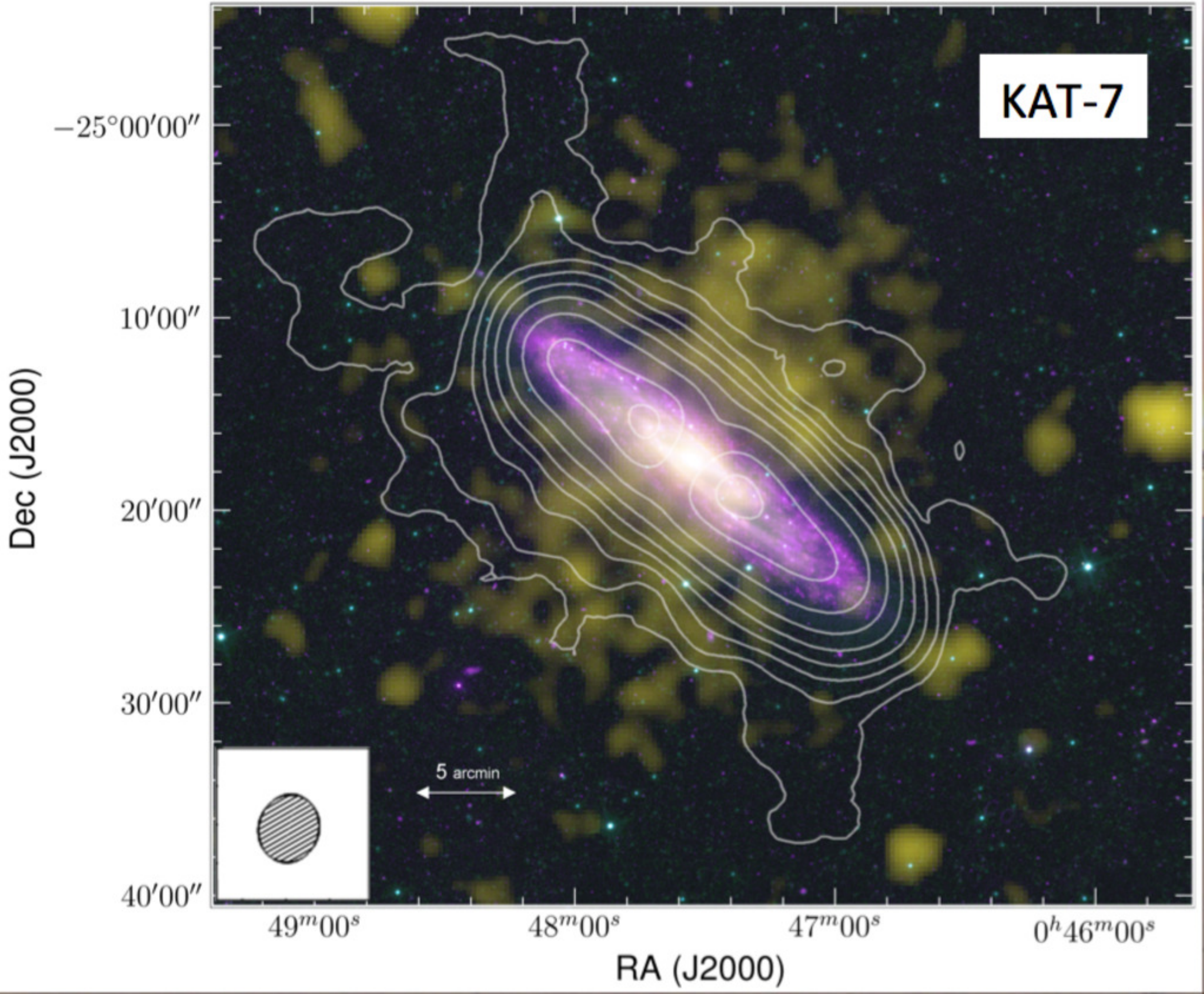}{fig:n253}{\hi\ distribution of NGC 253 (countours) superposed on a WISE image. The X-ray emission is in yellow.}

Besides individual nearby galaxies, the third most nearby cluster to us after Virgo and Fornax, Antlia, was also surveyed with KAT-7 \citep{hes15}. Those observations yielded 35 new \hi\ detections, 27 of which were new spectroscopic redshifts. The detected \hi\ allowed us to study the cluster substructure. Data from WSRT were also combined with KAT-7 data to study the southern filament of the Virgo cluster down to $8 \times 10^{17}$ cm$^{-2}$ \citep{sor15}. This allowed to detect the tail being stripped from NGC 4424 out to 60 kpc, three times longer than previously observed. More recently, M83 \citep{hea15} was also observed and KAT-7 recovered 5 times more flux than the VLA THINGS observations.

\section{Roll-out Plan for MeerKAT}

The South African MeerKAT radio telescope, an array of 64 dishes, is a precursor to the Square Kilometre Array (SKA) telescope and will be integrated into the mid-frequency component of SKA Phase 1 (SKA1-MID).  The  13.5 m  diameter dishes have an "offset gregorian" optical layout, meaning that no part of the surface is obstructed. 48 of the receptors are concentrated in the core area, which is approximately 1 km in diameter and the longest distance between any two receptors (the so-called maximum baseline) is 8 km. Here is a summary of the roll-out plan:
\begin{itemize}
\item {\bf March 2014:} the MeerKAT infrastructure was completed and commissioned. This included the roads, the antenna foundations, the assembly sheds and the new Karoo Array Processor Building (KAPS), which should also by used for SKA1-MID.
\item {\bf August 2015:} two antennas, populated with L-band receivers (sensitivity expected to be $\sim 320$m$^2$/K, rather than $\sim 220$m$^2$/K), become available to the project team. Intensive engineering tests will start in order to reduce risks prior to the full MeerKAT roll-out.
\item {\bf April 2016:} Array Release 1 (AR1) with 6 receptors, engineering verification completed. 
\item {\bf June 2016:} AR1 Science Commissioning completed. After AR1, the additional antennas get added on a rolling basis.
\item {\bf July 2016:} commissioning and science capability completed for a 16 antenna array (science on PI projects not scheduled to start yet).
\item {\bf Dec 2016:} engineering completed on AR2 (32 receptors).
\item {\bf April 2017:} AR2 science commissioning completed, early science (PI projects) starts, wideband imaging, pulsar timing.
\item {\bf Mid 2017:} engineering verification completed on AR3 (64 antennas).
\item {\bf End 2017:} AR3 Science Commissioning completed, start of the surveys as the additional science modes become available.
\end{itemize}

At this stage, observations will start for the 10 Large Survey Programs (LSP), which were allocated 5 years of observing time. Among those, the MHONGOOSE survey  will observe 30 galaxies for 200 hours each.
30\% of observing time will still be available for PI programs. See: http://public.ska.ac.za/meerkat/meerkat-schedule for updates on the schedule.

\section{Reaching the Cosmic Web with MeerKAT $+$ FAST}

Table \ref{cd} shows the sensitivities to low column density \hi\ of different telescope/observing time/resolution combinations for actual telescopes and for the future radiotelescope arrays.
With existing facilities,  column densities of $5.0 \times 10^{19}$ cm$^{-2}$ are reached with typical VLA (10 hours) and WSRT (12 hours) observations. With 10 times longer integrations, another factor of 10 can be gained with, for example WSRT \citep[HALOGAS:][]{hea11} or at lower spatial resolution ($\sim3.5$\arcmin) with KAT-7 \citep{car13, luc15, hes15}.
It was even possible to go down to $1.0 \times 10^{18}$ cm$^{-2}$ by combining KAT-7 and WSRT data. In 2017, MeerKAT should bring us to column densities $\sim 5.0 \times 10^{17}$ cm$^{-2}$ and we will have to wait the full SKA to reach the $10^{16}$ regime.

\begin{table}[!ht]
\caption{Expected sensitivities of different telescopes at 5$\sigma$.}
\label{cd}
\smallskip
\begin{center}
{\small
\begin{tabular}{lccccc}  
\tableline\tableline
\noalign{\smallskip}
Telescope & Integration & resolution & beam & sensitivity& Expected \\
Array(s) & hours & \kms\ & arcsecs & cm$^{-2}$ & date\\
\noalign{\smallskip}
\tableline
\noalign{\smallskip}
VLA (THINGS) & 10 & 5 & 30& $5.0 \times 10^{19}$ & \\
KAT-7 & 100 & 5 & 210 & $5.0 \times 10^{18}$ & \\
WSRT (HALOGAS) & 120 & 5 & 30 & $5.0 \times 10^{18}$ & \\
KAT-7 $+$ WSRT & 100 & 16 & 210 & $1.0 \times 10^{18}$ & \\
MeerKAT & 200 & 16 & 90 & $5.0 \times 10^{17}$ & 2017\\
SKA$_1$-MID & 100 & 5 & 30 & $7.5 \times 10^{17}$ & 2023 \\
SKA$_2$ & 10 & 5 & 30 & $2.5 \times 10^{17}$ & 2030\\
SKA$_2$ & 100 & 5 & 30 & $7.5 \times 10^{16}$ & 2030\\
\noalign{\smallskip}
\tableline\
\end{tabular}
}
\end{center}
\end{table}

\section{Conclusion}

However, in the near future (2017), the best combination to study low column density \hi\ with a good spatial resolution will be to combine the sensitivity of FAST with the spatial resolution of MeerKAT. The combination of the data from those two telescopes will allow, 6 years before SKA$_1$-MID, to do "cosmic web" research to levels $< 5.0 \times 10^{17}$ cm$^{-2}$, close to $10^{16}$ cm$^{-2}$, densities that would normally only be accessible to the full SKA around 2030. It is at those densities that we expect the galaxies to connect with the surrounding cosmic web.

\end{document}